\documentclass[twocolumn, prd,aps,tightenlines,notitlepage,nofootinbib,superscriptaddress,showpacs,preprintnumbers]{revtex4-1}
 \usepackage{amsmath,amssymb}
\usepackage{verbatim}
\usepackage{graphicx}
\usepackage[usenames]{color}
\usepackage{psfrag}
\usepackage{graphicx}
\usepackage{natbib} 
\usepackage{hyperref}
\usepackage{rotating} 
\usepackage{color}
\newcommand{\WMAP}{{\em WMAP}}
\newcommand{\COBE}{{\em COBE}}

\newcommand{\beq}{\begin{equation}}
\newcommand{\eeq}{\end{equation}}

\begin{document}

\title{ Fast analytic computation of cosmic string power spectra}

\author{Anastasios Avgoustidis} \email{anastasios.avgoustidis@nottingham.ac.uk}
\affiliation{School of Physics \& Astronomy\\
University of Nottingham,
Nottingham, NG7 2RD, England}

\author{Edmund J. Copeland} \email{ed.copeland@nottingham.ac.uk}
\affiliation{School of Physics \& Astronomy\\
University of Nottingham,
Nottingham, NG7 2RD, England}

\author{Adam Moss} \email{adam.moss@nottingham.ac.uk}
\affiliation{School of Physics \& Astronomy\\
University of Nottingham,
Nottingham, NG7 2RD, England}

\author{Dimitri Skliros} \email{dimitri.skliros@nottingham.ac.uk}
\affiliation{School of Physics \& Astronomy\\
University of Nottingham,
Nottingham, NG7 2RD, England}

\date{22 April 2013}

\begin{abstract}
We present analytic expressions for the cosmic string unequal time correlator (UETC) in the context of the Unconnected Segment Model (USM). This eliminates the need to simulate the many thousands of network realisations needed to estimate the UETC numerically. With our approach we can compute the UETC very accurately, over all scales of interest, in the order of $\sim 20-30$ seconds on a single CPU. Our formalism facilitates an efficient approach to computing Cosmic Microwave Background (CMB) anisotropies for cosmic strings. Discretising the UETC and performing an eigen-decomposition to act as sources in the {\tt CAMB} CMB code, the power spectrum can be calculated by summing over a finite number of eigenmodes. A much smaller number of eigenmodes are required compared to the conventional approach of averaging power spectra over a finite number of realisations of the string network. With the additional efficiency and performance improvements offered by the OpenMP {\tt CAMB} code, the time required to compute string power spectra is significantly reduced compared to the standard serial {\tt CMBACT} code. The latter takes $\sim 30$ hours on a modern single threaded CPU for 2000 network realisations. Similar percent level accuracy can be achieved with our approach on a moderately threaded CPU (8 threads) in only $\sim 15$ minutes. If accuracy is only required at the 10 percent level and the CPU is more highly threaded, cosmic string power spectra are now possible in  $ \sim 2-3$ minutes. This makes exploration of the string parameter space now possible  for  Markov-Chain-Monte-Carlo (MCMC) analysis.
\end{abstract}

\maketitle


\section{Introduction}

Results from the COsmic Background Explorer (\COBE)~\cite{Smoot:1992td} and Wilkinson Microwave Anisotropy Probe (\WMAP)~\cite{Spergel:2003cb} satellites have provided strong support that close to scale invariant fluctuations, with an amplitude of $\sim 10^{-5}$, were generated by a period of accelerated expansion (inflation) in the early Universe. This paradigm  provides an extremely good fit to other cosmological data, so by the principle of Occam's razor, having any additional source of fluctuations, such as topological defects, is deemed unnecessary. 

This line of reasoning, however, is only valid when inflation and defects are independent of each other. In many theories of the early Universe the two are intimately related. Cosmic strings, for example, are ubiquitous in models of supersymmetric~\cite{Linde:1993cn,Copeland:1994vg,Dvali:1994ms} and brane 
inflation~\cite{Dvali:1998pa,BMNQRZ,KKLMMT,BDKMcAll}. Inflation ends via a phase transition, and it is this symmetry breaking which results in strings~\cite{Jeannerot:2003qv,BMNQRZ,SarTye,CMP}. The amplitude of fluctuations created by strings can be comparable to the inflationary component, so it is crucial to account for them for a self-consistent theory.

However, strings should not be viewed as ``necessary evil" imposed on us by self-consistency.  Instead, they offer an exciting window of opportunity for exploring high-energy physics, through their potential observational effects. In particular, cosmic superstrings~\cite{Polch_Intro,CopKib} -- generically produced in string theoretic brane inflation models \cite{BMNQRZ,SarTye,DvalVil,CMP} -- form complicated multi-tension networks with segments joining together in Y-shaped junctions, and these characteristics depend quantitatively on fundamental string theory parameters and compactification data. Thus, they offer one of the very few observational windows into string theory, and even the non-detection of strings can constrain fundamental string theory parameters~\cite{ACMPPS}.  For the simplest networks of ``ordinary" field theory strings, non-detection results in a constraint on the energy scale of the string-producing phase transition, which marks the end of the inflationary phase. For recent reviews of the opportunities offered by cosmic strings for probing inflationary models and string theory, see~\cite{CopPogVach,Hindm_Rev}.      

In this paper we consider the observational signature of cosmic strings on the Cosmic Microwave Background (CMB). Strings create anisotropies by their motion along the line-of-sight (the Kaiser-Stebbins effect~\cite{Kaiser:1984iv}) between ourselves and the last-scattering surface, as well as other related physics at the epoch of recombination. Since the formation of strings is a non-Gaussian process, there are associated non-Gaussian signatures in the CMB (see Ref.~\cite{Ringeval:2010ca} for a recent review). In this work, however, we only consider the Gaussian component, that is the CMB power spectrum. 

The amplitude of string fluctuations is typically characterised by their dimensionless mass per unit length, $G\mu/ c^2$. Current CMB power spectra allow for $G\mu/c^2 \lesssim 2-6 \times 10^{-7}$ at $ 2 \sigma$ confidence~\cite{Battye:2010xz,Urrestilla:2011gr,Dvorkin:2011aj}. The range in this limit arises primarily due to two reasons: (1) the precise details of the evolution of the string network, i.e. its scaling properties; (2) modelling uncertainties due to the dynamical range of simulations. A recent discussion of the various assumptions and problems with each method can be found in Ref.~\cite{Battye:2010xz}.
 
No matter which method is used, the key intermediate quantity of interest is the Unequal Time Correlator (UETC) of the string network~\cite{TurPenSelj}. There exists a separate UETC for each of the scalar, vector and tensor contributions to the energy-momentum tensor. From this, one can then calculate CMB anisotropies using an Einstein-Boltzmann code. The UETC can be estimated directly from simulations. An alternative method is to model the strings as an ensemble of uncorrelated straight segments, with a root-mean-square (RMS) velocity and correlation length chosen to match that of simulations. The latter approach is known as the Unconnected Segment Model (USM)~\cite{Albrecht:1997mz,cmbact}, and by modelling strings as 1D objects it avoids many of the issues associated with dynamical range. It is also easier to vary the network parameters within the context of the model and check their impact on observable quantities. The USM has disadvantages of its own however, which are discussed in Ref.~\cite{Battye:2010xz}.

Up to now the UETC of the USM has only been (indirectly) calculated numerically, by simulating an ensemble of source histories. The physics of the USM is relatively simple, however, and in this paper we show the UETC can be derived analytically. This is an important result for two reasons. The first is that, since the the UETC encodes all information of the string model, it is extremely useful to be able to evaluate this quickly and assess the impact of changing parameters, or to compare it with full numerical simulations of the string network.  The second is that knowledge of the UETC allows the CMB power spectrum to be computed in a more efficient way, rather than averaging over source histories of the USM.

Our formalism has particular relevance to the next generation of experiments which will measure the polarisation signal of the CMB. Since strings generate vector and tensor fluctuations, they also create a B-mode polarisation signal. In many theories of the early Universe which give rise to inflation {\em and} strings, the B-mode signal from inflation can be much {\em smaller} than that from strings. This means the string B-mode signal could provide a unique window to testing high-energy physics at the $\sim 10^{16}$ GeV scale. In a future publication we will use the analytic form of the UETC to approximate, analytically, the string B-mode spectrum in terms of the underlying parameters.

The outline of the paper is as follows. In section~\ref{sec:uetc} we derive analytically  the cosmic string UETC's. In section~\ref{sec:cmb} we 
use these as inputs to an Einstein-Boltzmann code to compute CMB temperature and polarisation power spectra. 
Our method dramatically improves computational time, making exploration of the string parameter space 
now possible  for  Markov-Chain-Monte-Carlo (MCMC) analysis.  Finally we conclude and outline future directions of work. Throughout this paper we will work in relativistic units with $c=1$.


\section{String unequal time correlator} \label{sec:uetc}

Perturbations generated by cosmic strings are qualitatively different from those arising 
from inflationary vacuum fluctuations.  While inflationary perturbations are \emph{passive}, in that 
they are set as initial conditions (on super-horizon scales) and are then evolved through the 
``source free'' linearised Einstein-Boltzmann equations, cosmic strings generate fluctuations 
which add a forcing term to these equations.  Thus, strings \emph{actively} source 
metric perturbations (on sub-horizon scales) at all times, and so their cosmological evolution is of 
key importance for predicting CMB anisotropies.  Indeed, CMB spectra can be computed 
from the unequal time correlators (UETC's) of the network energy-momentum tensor components 
(in Fourier space), which are manifestly functions of the time-dependent network 
parameters.  

In this section, we describe our {\em analytic} approach for computing the relevant UETC's for 
cosmic string networks.  Starting from the (Fourier transform of the) energy-momentum tensor 
of a single string segment, we review and update the Unconnected Segment Model 
(USM) \cite{Albrecht:1997mz,cmbact} procedure for building the energy-momentum tensor 
components for a network of strings.  Given the
UETC's, these are then used as inputs to an Einstein-Boltzmann code to calculate CMB power spectra. This is a key step in improving the speed with which we can calculate spectra over existing string codes.

The fact that strings are active sources allows them to produce significant vector 
perturbations, which normally decay in passive models where they are not continuously 
sourced.  In fact, it is the vector metric perturbation that provides the dominant contribution 
to the B-mode spectrum, which is of particular interest for future CMB polarisation experiments.  Unlike inflation, where 
tensor modes are suppressed by a factor proportional to the slow-roll parameter, strings 
generally also produce a non-negligible amount of tensor perturbations.  We therefore need to consider 
all three contributions, scalar (S), vector (V) and tensor (T), and compute all the relevant UETC's.  


\subsection{Evolving string networks}

Since cosmic strings are active seeds of perturbations, their cosmological evolution affects the CMB. 
In order to confront cosmic strings with CMB data, it is therefore of crucial importance 
to be able to accurately model the behaviour of string networks throughout cosmic 
history. 

Over the last couple of decades, significant progress has been made in this direction 
and a number of complementary techniques have been developed to describe the 
evolution of a wide range of string models (For reviews see 
Refs.~\cite{book,Hindmarsh:1994re,CopKib,Hindm_Rev}). 
The dynamics of ``ordinary" string defects is fully described by non-linear field theories in cosmological backgrounds, 
which can be evolved numerically in lattice simulations~\cite{Vincent:1997cx,Moore:2001px,Bevis:2006mj}. 
For cosmological applications, short scale effects  are generally neglected (even though there remains 
a question as to the validity of this assumption, see for example Ref.~\cite{Hindmarsh:2008dw})  
and the networks can be approximated by a zero-width, Nambu-Goto string limit. With some additional 
input from field theory to 
describe the result of string inter-commutations~\cite{Shell_Recon,Matzner}, this has allowed 
high-resolution simulations of Nambu-Goto strings in realistic 
cosmologies~\cite{Bennett:1989yp,Allen:1990tv,RingSakBouch,MartShell,B-POS}. 
Comparing numerical results with analytic approaches, it has also been realised that key 
features of the complicated network dynamics can be captured by surprisingly simple 
analytic models, describing the string network by only a handful of macroscopic variables.  

In this approach, one starts with the assumption (justified by simulations) that the 
string network has a random walk structure, so it can be characterised -- on large 
scales -- by a correlation length, $L$, quantifying the average length of string segments   
and the average separation between them\footnote{For a random walk, these two 
length-scales effectively coincide and in simulations of the simplest models of strings they 
are found to be in good numerical agreement.}.  Similarly, string motion can be described 
by a root-mean-square (RMS) velocity, $v$, of string segments. These two variables satisfy  
macroscopic equations of motion, derived directly from the Nambu-Goto action, after 
averaging over the worldsheet and introducing a phenomenological interaction term~\cite{Kibble,Austin:1993rg,VOS,VOSk}
 \begin{eqnarray}\label{Ldt}
   \frac{1}{L}\frac{dL}{dt} & =& (1+v^2)H + \frac{\tilde c v}{2L}\,, \\
   \label{vdt}
   \frac{dv}{dt} &=& (1-v^2)\left(\frac{\tilde k}{L}-2Hv\right)\,,
 \end{eqnarray}  
where $H$ is the Hubble function, $H=\dot{a}/a$, $a(t)$ is the scale factor and an overdot represents the derivative with respect to physical time $t$. The parameter $\tilde c$ is a 
constant quantifying energy loss due to loop production and can be calibrated 
by comparison to simulations. Finally, the curvature parameter ${\tilde k}$ 
indirectly encodes information about the small-scale structure on strings 
and can be expressed in terms of $v$. For relativistic strings we have~\cite{VOSk}
 \begin{eqnarray}\label{keqn}
{\tilde k}=\frac{2\sqrt{2}}{\pi}\left(\frac{1-8v^6}{1+8v^6}\right)\,.
\end{eqnarray}     
Equations (\ref{Ldt}-\ref{keqn}) form the Velocity-dependent One-Scale (VOS) model, 
and have been shown to be in remarkable agreement with Nambu-Goto and field 
theory simulations for the simplest of cosmic string models. Writing the correlation 
length in terms of a new variable $\epsilon$,
\begin{equation}
\hspace{3.5cm} L=\epsilon t, 
\end{equation}
and expressing the Hubble function in 
terms of the expansion parameter $\beta$, such that $a(t)\propto t^\beta$, one 
finds that this system has the scaling solution
 \begin{eqnarray}
\epsilon&=&\sqrt{ \frac{{\tilde k}({\tilde k}+\tilde{c})}{4\beta(1-\beta)} }\,,
\label{vosscaling1}
\\
v&=&\sqrt{ \frac{{\tilde k}(1-\beta)}{\beta({\tilde k}+\tilde{c})} } \,,
\label{vosscaling2}
 \end{eqnarray}
which is the generic attractor.  In comoving units, the correlation length is then
\begin{equation}
l = \frac{L}{a}=\xi \tau \,, 
\end{equation}
where $\tau$ is the comoving horizon (or alternatively conformal time).  Since, in a given era ($\beta=\rm const$, equal to 1/2 and 2/3 in the radiation and matter eras, respectively), the 
physical horizon is $d_H=a\tau=t/(1-\beta)$, the relation between $\epsilon$ and 
$\xi$ is
\begin{equation}
\xi=\epsilon (1-\beta) \,.
\end{equation}

For string networks of higher complication, more sophisticated VOS models can be constructed 
which include additional parameters. For example, superconducting strings~\cite{Witten:1984eb} carry 
currents described by a charge parameter~\cite{Oliveira:2012nj}, while cosmic 
superstrings~\cite{Polch_Intro,CMP} can be modelled as a collection of several network components, each 
with a different correlation length, velocity and loop production efficiency, as well as several extra interaction 
coefficients describing the formation of string junctions~\cite{JackJoPolch,TWW,NAVOS,Avgoustidis:2009ke}.  
The generalisation of our methods to such networks will be the subject of a follow-up publication.  

Here, we will consider one extra parameter, $\alpha$, describing the effect of short-scale 
wiggles on the string energy and tension, thus affecting the UETC.  To a first approximation, 
the VOS equations do not depend directly on $\alpha$, as the main effects of short-scale-structure
are already captured by ${\tilde k}$ in equation (\ref{keqn}). More sophisticated ``wiggly" models exist~\cite{wiggly}, 
but the VOS equations above are sufficient to reproduce simulation results with respect 
to string correlation lengths and velocities.  The parameter $\alpha$ can be readily 
obtained by Nambu-Goto simulations. Note that, physically, $\alpha$ and ${\tilde k}$ must be 
implicitly related, but a quantitative formula is not known.         

The easiest way to describe the string wiggliness $\alpha$ is by writing the
string energy-momentum tensor as
\begin{widetext}
\begin{equation}\label{Tmunu}
\Theta^{\mu\nu}(y)=\frac{1}{\sqrt{-g}} \int d\tau d\sigma  
\left[U \sqrt{\frac{-x'^2}{\dot x^2}}\dot{x}^\mu \dot{x}^\nu - 
T \sqrt{\frac{-\dot x^2}{x'^2}} x'^{\mu} x'^{\nu}\right]
\delta^{(4)} ( y -  x(\sigma ,\tau))\,, 
\end{equation}  
\end{widetext} 
where $U$ is the string energy per unit length and $T$ the string 
tension.  Lorentz invariance requires $U=T\equiv \mu$.  However, 
if short scale wiggles cannot be resolved, this has the effect of 
increasing the string energy per unit length and reducing the 
tension~\cite{Vilenkin_ss, Carter_wiggly} in such a way that their product 
remains constant\footnote{This is also the case if Lorentz invariance is broken 
at a fundamental level, e.g. in superconducting strings~\cite{Carter_super}.}
\begin{equation}
UT=\mu^2 \,.
\end{equation}
Then, an effective ``coarse-grained" energy-momentum tensor can be written 
in the form of  (\ref{Tmunu}) with ``renormalised" string energy density and tension, 
parameterised through $\alpha$ as follows
\begin{equation}
U=\alpha\mu \ , \ T=\frac{\mu}{\alpha} \,.
\end{equation} 
As we will see, this parameter $\alpha$ enters directly into the UETC through this 
effective energy-momentum tensor for coarse-grained string segments. 


\subsection{Energy-momentum tensor from a single string}

Let us consider a single straight segment of the string network. We are free to orientate the wave vector as ${\bf k} = k \hat{k}_3$, where $k$ is its magnitude, in which case the real component of the Fourier transform of the energy-momentum tensor (\ref{Tmunu})  becomes (e.g.~\cite{Albrecht:1997mz,cmbact}) 
{\begin{eqnarray}
\Theta_{00} &=& \frac{\mu \alpha}{\sqrt{1-v^2}} \frac{\sin(k \hat{X}_3 \xi \tau/2)}{k \hat{X}_3 /2} \cos \left(\chi + k \hat{\dot{X}}_3 v \tau \right)\,, \\
\Theta_{ij} &=& \left[v^2 \hat{\dot{X}}_i \hat{\dot{X}}_j - \frac{1-v^2}{\alpha}\hat{X}_i \hat{X}_j \right] \Theta_{00}\,,
\end{eqnarray}
where $\hat{X}$ and $\hat{\dot{X}}$ are randomly orientated unit vectors satisfying $\hat{X} \cdot \hat{\dot{X}} = 0$ 
and $i,j=1\dots 3$. The phase of the Fourier mode is set by the location of the string, ${\bf x}_0$, where $\chi = {\bf k} \cdot {\bf x}_0$. One can then identify the scalar, vector and tensor anisotropic stress by 
\begin{eqnarray}
\Theta^S &=& \left(2 \Theta_{33} - \Theta_{11} - \Theta_{22} \right)/2\,, \\ 
\Theta^V &=& \Theta_{13}\,,\\
\Theta^T &=& \Theta_{12}\,.
\end{eqnarray}
For each string segment this gives 
\begin{eqnarray}
\frac{2 \Theta^S}{\Theta_{00}} &=&  v^2 \left(3\hat{\dot{X}}_3 \hat{\dot{X}}_3-1 \right) - \frac{1-v^2}{\alpha^2} \left(3 \hat{{X}}_3 \hat{{X}}_3 -1 \right), \\
\frac{\Theta^V}{\Theta_{00}} &=&  v^2 \hat{\dot{X}}_1 \hat{\dot{X}}_3 - \frac{1-v^2}{\alpha^2}  \hat{{X}}_1 \hat{{X}}_3  \,, \\
\frac{\Theta^T}{\Theta_{00}} &=&  v^2 \hat{\dot{X}}_1 \hat{\dot{X}}_2 - \frac{1-v^2}{\alpha^2}  \hat{{X}}_1 \hat{{X}}_2  \,.
\end{eqnarray}
We choose $\hat{{X}}$ to be the usual position vector in spherical coordinates, i.e. $\hat{{X}}^T= \left(\sin \theta \cos \phi, \sin\theta \sin\phi, \cos\theta \right)$. Given the orthogonality of $\hat{{X}}$  and $\hat{\dot{X}}$, once $\hat{{X}}$ is specified there is an angle $\psi$ from 0 to $2 \pi$ which gives $\hat{\dot{X}}$,
\beq
\hat{\dot{X}} = \left[ \begin{array}{c}   \cos \theta \cos \phi \cos \psi - \sin \phi \sin \psi \\  \cos \theta \sin \phi \cos \psi + \cos \phi \sin \psi \\ - \sin \theta \cos \psi \end{array}\right]\,.
\eeq
We can then generalise to a network of strings comprising many segments with different 
orientations, by averaging over angles.

\subsection{Unconnected segment model}

The Unconnected Segment Model (USM) framework~\cite{Albrecht:1997mz,cmbact} models the string network as a set of uncorrelated straight segments, each moving with random velocity. All segments are assumed to be produced at some fixed initial time. Throughout cosmic history a certain fraction of these segments decay at each epoch in order to maintain scaling of the network. Since, during scaling, the number density of strings falls as $n(\tau) \propto \tau^{-3}$, one needs to track an extremely large number of segments in order to have of order one segment remaining today.  

To avoid tracking each segment, the USM consolidates all string segments that decay at the same discretised epoch $\tau_i$ into a single string.  Specifically, the number of segments which decay between conformal time $\tau_{i-1}$ and 
$\tau_i$ is
\beq
N_d (\tau_i) = V \left[ n(\tau_{i-1}) - n(\tau_i) \right]\,,
\eeq
where $V$ is the simulation volume and $n(\tau)$ the number density of strings, given by
\beq \label{eqn:ctaucoeff}
n(\tau) = \frac{C(\tau)}{(\xi \tau)^3}\,.
\eeq
The coefficient $C(\tau)$ is determined by requiring the total number of strings at any time be equal to $V/(\xi \tau)^3$. 
In Ref.~\cite{cmbact} it was found to be approximately constant and of order unity throughout the simulations
performed.

The $N_d (\tau_i)$ string segments are then consolidated into a single segment, which has an energy-momentum weight $\sqrt{N_d (\tau_i)}$. The total energy-momentum tensor is a sum of these consolidated segments, given by
\beq \label{eqn:em_consol}
\Theta_{\mu \nu} = \sum_{i=1}^K \left[N_d (\tau_i) \right]^{1/2} \Theta_{\mu \nu}^i T^{\rm off} (\tau,\tau_i,L_f)\,,
\eeq
where $\Theta_{\mu \nu}^i$ is the energy-momentum tensor of a single segment $i$, $K$ is the number of consolidated 
segments and $T^{\rm off} (\tau,\tau_i,L_{f}) $ is a function that controls the rate of string decay. This has the form 

\beq
T^{\rm off} (\tau,\tau_i,L_{f}) = \left\{ \begin{array}{rl}
 1 & \quad \quad \mbox{$\tau<L_{f} \tau_i$}\,, \\
\frac{1}{2} + \frac{1}{4} \left(y^3 - 3y \right) & \quad \quad \mbox{$L_{f} \tau_i<\tau<\tau_i$}\,, \\
  0 &\quad \quad \mbox{$\tau>\tau_i$}\,,
       \end{array} \right.
\eeq
where 
\beq
y = 2 \frac{\ln (L_{f} \tau_i/\tau)}{\ln(L_{f})}-1\,.
\eeq
Segments start to decay at $L_f \tau_i$ and have disappeared completely at $\tau_i$. The parameter $L_f<1$ therefore controls how fast the segments decay, and approximates the Heaviside step function when $L_f \rightarrow 1$. The value of this parameter is less well-understood from Nambu-Goto simulations. In general, higher $L_f$ results in less power in the CMB power spectrum~\cite{Albrecht:1997mz,cmbact} for a fixed string tension $G\mu$. 


\subsection{Analytic expressions}

To compute the unequal time correlator (UETC) analytically we integrate over all strings in the network
\begin{widetext}
\beq
\langle \Theta(k, \tau_1) \Theta(k,\tau_2) \rangle = \frac{2 f(\tau_1,\tau_2,\xi,L_f)}{16\pi^3} \int_0^{2 \pi} d \phi \int_0^{\pi} \sin \theta \, d \theta \int_0^{2 \pi} d \psi \int_0^{2 \pi} d \chi \, \Theta(k,\tau_1) \Theta(k,\tau_2) \,,
\eeq
\end{widetext}
where by $\Theta$ we mean $\Theta_{00}, \Theta^S, \Theta^V, \Theta^T$ and $f(\tau_1,\tau_2,\xi,L_f)$ is a scaling factor associated with the decrease in the number density and the decay of strings (see below). The factor of two in the numerator arises from us only considering the real part of the UETC.  Three of the integrals can be performed to give the remaining $\theta$ integral in terms of Bessel functions. Some of the resulting terms do not have compact analytic expressions, so we need to make use of series expansions.  In each case we can write the remaining 
$\theta$ integral as the sum
\begin{eqnarray} \label{eqn:acoeff}
&& \langle \Theta(k, \tau_1) \Theta(k,\tau_2) \rangle = \frac{  f(\tau_1,\tau_2,\xi,L_f) \mu ^2}{k^2 \left(1-v^2\right)} \times  \\ \nonumber && \sum_{i=1}^6 A_i \left[ I_i (x_-, \rho) - I_i (x_+, \rho) \right] \,,
\end{eqnarray} 
where the relevant integral identities $I_i (x_{\pm},\rho)$ are given in Appendix~\ref{app:int},  $\rho = k |\tau_1 - \tau_2| v$, $x_{1, 2} = k \xi \tau_{1, 2}$ and $x_{\pm} = (x_1 \pm x_2 )/2$. For each scalar, vector and tensor UETC we write down the amplitude $A_i$ of each integral component, in Appendix~\ref{app:int}.

We will also be interested in certain asymptotic limits of the UETC. For completeness we write down the small $x$ limit,
\begin{equation} \label{eqn:bcoeff}
\langle \Theta(k, \tau_1) \Theta(k,\tau_2) \rangle \approx \frac{  f(\tau_1,\tau_2,\xi,L_f) \mu ^2}{k^2 \left(1-v^2\right)} B \,.
\end{equation}
Finally, we are also interested in the equal time correlator (ETC), so that $x_1 = x_2 = x_{+} = x$ and $x_-=\rho=0$. In this case we write the exact expression
\begin{equation} \label{eqn:ccoeff}
\langle \Theta(k, \tau) \Theta(k,\tau) \rangle = \frac{  f(\tau,\tau,\xi,L_f) \mu ^2}{k^2 \left(1-v^2\right)} C \,.
\end{equation}
The $B$ and $C$ coefficients are also given in Appendix~\ref{app:int}. Equations~(\ref{eqn:acoeff}),~(\ref{eqn:bcoeff}) and~(\ref{eqn:ccoeff}) form the basis of our fast UETC code.


\subsection{Scaling factor}

In the above we introduced the scaling factor $f(\tau_1,\tau_2,\xi,L_f)$. If we assume the network consists of only a single string whose number density per simulation volume is fixed and the string does not decay, then from (\ref{eqn:em_consol}) in the USM $\Theta_{\mu \nu}=\Theta_{\mu \nu}^1$, while analytically we can set $f(\tau_1, \tau_2,\xi,L_f) = 1$. Next, we turn on the scaling and decay of strings. This means in the USM the UETC is given by 
\begin{widetext}
\beq
\langle \Theta_{\mu \nu} (k,\tau_1)\Theta_{\rho \sigma} (k,\tau_2) \rangle= \langle \Theta_{\mu \nu}^1 (k,\tau_1) \Theta_{\rho \sigma}^1 (k,\tau_2) \rangle  \sum_{i=1}^K N_d (\tau_i) T^{\rm off} (\tau_1,\tau_i,L_f) T^{\rm off} (\tau_2,\tau_i,L_f) \,,
\eeq
\end{widetext}
since $\langle \Theta^i \Theta^j \rangle$ is equal to $\langle \Theta^1 \Theta^1 \rangle$ for $i=j$ and 0 for $i \ne j$. Increasing the number of segments such that $K \rightarrow \infty$ then  the sum when $L_f \rightarrow 1$ can be evaluated to give $C(\tau) \rightarrow 1$ (defined in Eqn.~\ref{eqn:ctaucoeff}). This gives the scaling factor 
\beq \label{eqn:scale_L1}
f(\tau_1,\tau_2,\xi,L_f \rightarrow 1) = \frac{1}{\left[\xi \, {\rm Max}(\tau_1,\tau_2) \right]^3}\,.
\eeq
The $0<L_f<1$ scaling factor is much more lengthy but is possible to write down analytically. For the purposes of our work we assume the strings segments decay instantaneously and set $L_f \rightarrow 1$.


\subsection{Comparison with simulations}

The expressions for the UETC's given in Appendix~\ref{app:int} have been coded in a self-contained Fortran 90 module. In this code, we make use of asymptotic limits to improve speed and accuracy. For small $x_1$ and $x_2$, we use the small $x$ expansion (the $B$ coefficients), thereby eliminating the need to evaluate trigonometric functions or perform a series expansion of spherical Bessel functions. For $x_1 \approx x_2$, we use the form of the ETC (the $C$ coefficients), as the amplitudes $A_i$ contain terms $\propto 1/\rho^2$, which diverge when $x_1=x_2$ (this divergence in the $A_i$ coefficients vanishes when fully expanding out the integral in this limit). 

In all other cases we perform the sum of integral components, tuning the number of terms in the spherical Bessel series expansions which are required for sufficient accuracy, given $x_1$ and $x_2$ (in general larger $x_1$ and $x_2$ require more terms). A slight caveat here is that the individual terms of the series expansion can become much larger than the total sum, and for $x_1$ or $x_2$ greater than $\sim 30$ loss of accuracy can occur due to the limitation of double precision arithmetic in our code. In these cases we perform the integrals of (\ref{eqn:j0_series}) and (\ref{eqn:j1_series}) numerically. The result is a code which can generate the scalar, vector and tensor UETC's, over all scales of interest relevant for the CMB, on a single CPU in $\sim 20-30$ seconds. We have parallelised this code, so a moderately threaded CPU can calculate the UETC's extremely quickly. 

For our comparisons with the (numerical) USM we use the publicly available {\tt CMBACT} code~\cite{cmbact}, using a string decay parameter of $L_f=0.99$ and $K=500$ consolidated segments. We assume this value of $L_f$ is close enough to unity so Eqn.~(\ref{eqn:scale_L1}) is valid in the analytic calculation. We use string network parameters $v=0.65$, $\xi=0.13$ and $\alpha = 1.9$, turning off the time evolution of these in {\tt CMBACT}. These parameters are chosen to match the VOS  evolution in the radiation era, as in Refs.~\cite{Battye:1997hu,cmbact}.  At present, we have not attempted to incorporate time dependence of the string parameters in our code, as this complicates the eigen-decomposition of the UETC (see below).  Although the scaling solution will vary somewhat as the Universe changes from radiation to matter domination, for the purposes of our model we fix the network parameters to be constant. 

\begin{figure}
\centering
\mbox{\resizebox{0.238\textwidth}{!}{\includegraphics[angle=0]{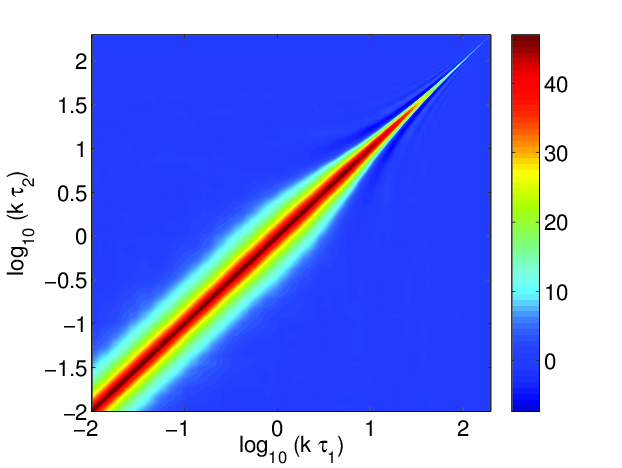}}}
\mbox{\resizebox{0.238\textwidth}{!}{\includegraphics[angle=0]{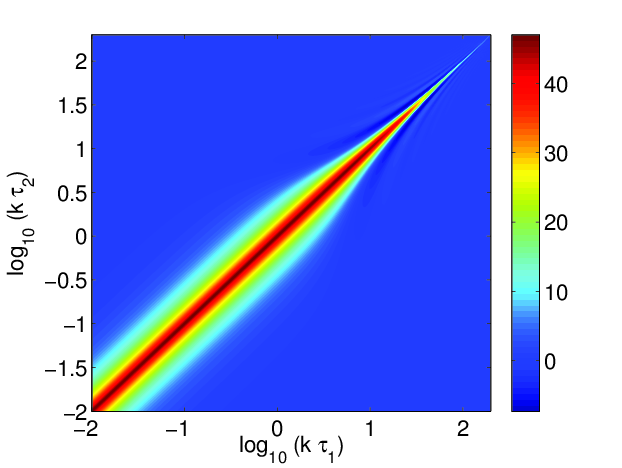}}}

\mbox{\resizebox{0.238\textwidth}{!}{\includegraphics[angle=0]{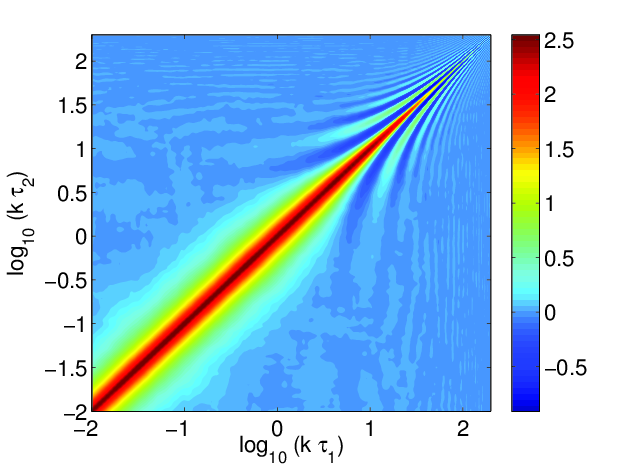}}}
\mbox{\resizebox{0.238\textwidth}{!}{\includegraphics[angle=0]{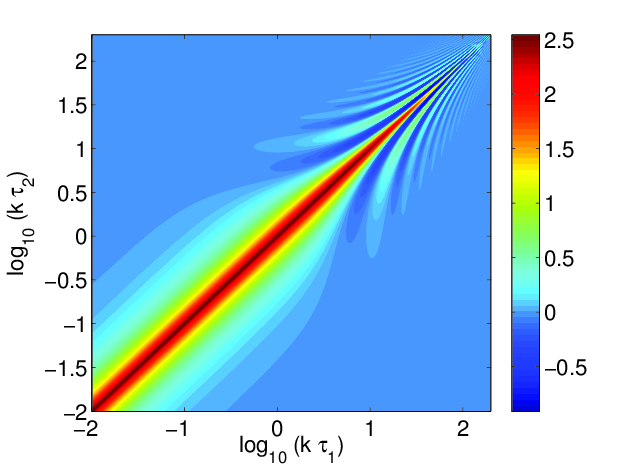}}}

\mbox{\resizebox{0.238\textwidth}{!}{\includegraphics[angle=0]{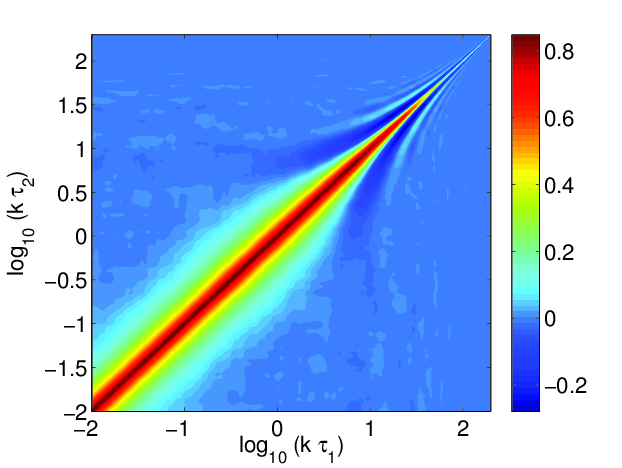}}}
\mbox{\resizebox{0.238\textwidth}{!}{\includegraphics[angle=0]{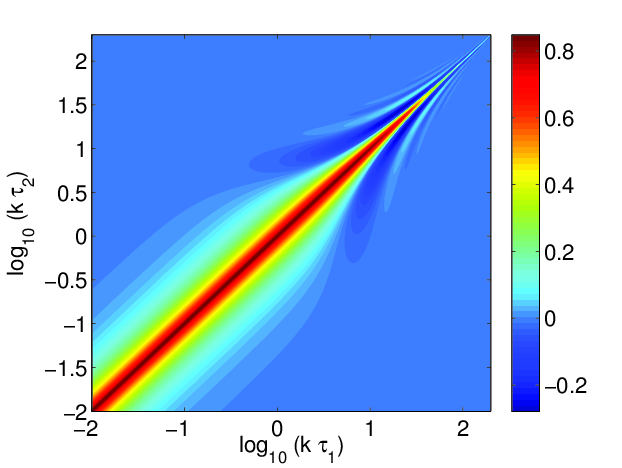}}}

\mbox{\resizebox{0.238\textwidth}{!}{\includegraphics[angle=0]{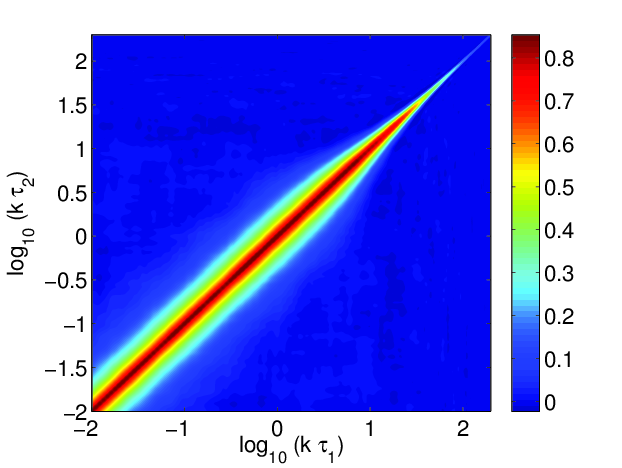}}}
\mbox{\resizebox{0.238\textwidth}{!}{\includegraphics[angle=0]{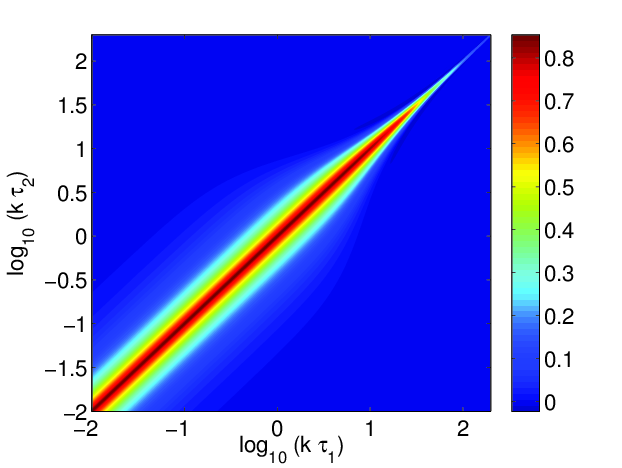}}}

\mbox{\resizebox{0.238\textwidth}{!}{\includegraphics[angle=0]{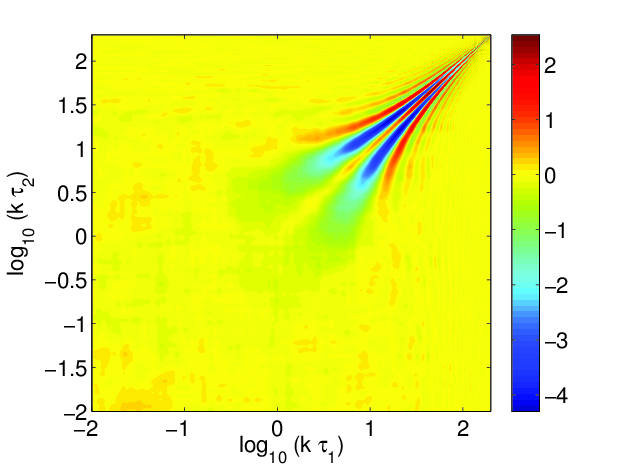}}}
\mbox{\resizebox{0.238\textwidth}{!}{\includegraphics[angle=0]{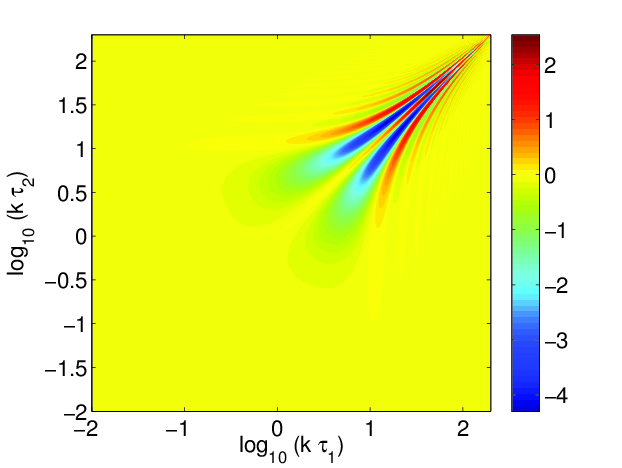}}}

\caption{\label{fig:uetc_scale} Scaled unequal time correlators (UETC's), $\left( \tau_1 \tau_2 \right)^{1/2}  \langle \Theta (k,\tau_1) \Theta (k,\tau_2) \rangle /\mu^2$. (Left) The UETC is derived from 20000 realisations of the string network using {\tt CMBACT}, with a decay parameter $L_f=0.99$, and (right) the analytic calculation (see text). From top-to-bottom we show the 00 component, then scalar, vector and tensor anisotropic stress components, and finally the $00-S$ cross correlator.} 
\end{figure}

In Fig.~\ref{fig:uetc_scale} we show {\em scaled} UETC's,  defined as $\left( \tau_1 \tau_2 \right)^{1/2}  \langle \Theta (k,\tau_1) \Theta (k,\tau_2) \rangle /\mu^2$, with the numerical and analytical results shown respectively in the left and right columns. From our analytic expressions  for the UETC one can see that the only dynamical variable is the combination $k \tau$, which is in agreement with physical expectations. We plot the UETC derived from both 20000 realisations of {\tt CMBACT} and our analytic expressions, and find the two in excellent agreement. For comparison, deriving the UETC from 20000 realisations of {\tt CMBACT} takes several hours\footnote{This does not include the Einstein-Boltzmann solver part of {\tt CMBACT}, which takes much longer to run.}.


\section{CMB anisotropies} \label{sec:cmb}

An important source of difficulty in the computation of CMB power spectra from cosmic strings is that they are \emph{incoherent} sources of perturbations.  As a result one faces complications in calculating the ensemble average of the CMB transfer functions.  In general there are two approaches to address the issue of incoherence.  The first avoids dealing with the UETC directly. Instead, one creates an ensemble of source histories of the USM, which has the same 2-point correlation statistics as the UETC. The source histories are then used in an Einstein-Boltzmann code, and the power spectrum can be obtained by averaging over many realisations of the network. This is the approach {\tt CMBACT} uses. 

The second works directly with the UETC, which can be estimated from string simulations, or in our case analytically. The UETC is first decomposed into eigenmodes. Each of the individual modes is {\em coherent}, and can be used as source function in the Einstein-Boltzmann code. This approach was originally proposed in Ref. ~\cite{Pen:1997ae} and is the method used by groups calculating CMB spectra from Abelian-Higgs string simulations (e.g.~\cite{Bevis:2006mj}). Since we readily have the UETC, we follow the latter approach.

\subsection{Eigenmode decomposition} 

Since the scaled UETC is {\em only} a function of $k \tau_{1,2}$ we can discretise it on a logarithmic grid in $k \tau_{1,2}$ with $n \times n$ grid points. It is also real and symmetric and hence diagonalisable. We therefore decompose it as a sum of eigenvectors 
\begin{eqnarray} \label{eqn:eigen}
&& \left( k^2  \tau_1 \tau_2 \right)^{\gamma} \left( \tau_1 \tau_2 \right)^{1/2}  \langle \Theta (k,\tau_1) \Theta (k,\tau_2) \rangle = \\ \nonumber && \sum_{i=1}^n \lambda_i u_i (k \tau_1) \otimes u_i (k \tau_2)\,, 
\end{eqnarray}
where $\lambda_i$ are the eigenvalues, and the eigenvectors $u_i$ are also only functions of $k \tau$. The term $\left( k^2  \tau_1 \tau_2 \right)^{\gamma} $ is a weighting factor, designed to improve the accuracy of the reconstructed UETC when the eigenmode sum is truncated at some value less than $n$ (see below). A weight of $\gamma>0$, for example, will improve the reconstruction of the $k \tau > 1$ region. In practice this value is tuned based on the scales which contribute to CMB anisotropies. We find a weight of $\gamma = 0.25$ is around optimal for improving the convergence of the power spectra.

Since the correlation between scalars, vectors and tensors vanishes we can perform the diagonalisation on the vector/tensor UETC's independently. However, for the scalars we have two components of the energy-momentum tensor, the density, $\Theta_{00}$, and anisotropic stress, $\Theta^S$, which are correlated. In order to perform the decomposition for scalars we therefore  discretise the UETC on a $2n \times 2n$ grid. This has a $2 \times 2$ block diagonal form, with the  $00-00$ and $S-S$ components of~(\ref{eqn:eigen}) along the diagonal, and $00-S$ as the off-diagonal components. After performing the eigen-decomposition the first half of the eigenvector corresponds to density, and the other half to the anisotropic stress. 

This diagonalisation procedure introduces a change of basis. Since the eigenvectors are orthogonal, and the Einstein-Boltzmann solver uses {\em linear} perturbation theory, one can use each of these new basis functions as a source term. This means we can simply modify the Einstein equation sources according to
\beq
\Theta (k \tau) \rightarrow \frac{u(k \tau)} {(k \tau)^\gamma \tau^{1/2} }\,.
\eeq
The total power spectrum is then found  by summing over all eigenmodes
\beq
C_{\ell} = \sum_{i=1}^n \lambda_i C_{\ell}^i\,,
\eeq
where $C_{\ell}^i$ are the individual power spectra of each eigenmode. In practice we order modes from the highest to lowest eigenvalues and truncate the sum at a finite number of modes (which gives the desired accuracy for $C_{\ell}$).  Although each of the eigenmodes is coherent, good convergence can be obtained including only relatively few modes in the sum.


\subsection{CMB power spectrum}

\begin{figure*}
\centering
\mbox{\resizebox{0.99\textwidth}{!}{\includegraphics[angle=0]{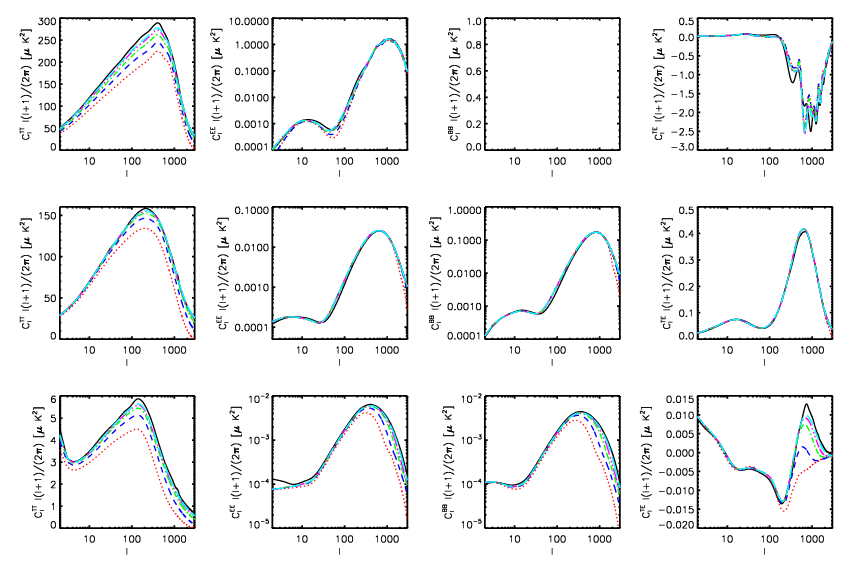}}}
\caption{\label{fig:uetc_cls} Comparison with {\tt CMBACT}. On the top panel we show scalars, in the middle vectors and on the bottom tensors. From left-to-right we show the TT, EE, BB and TE power spectra. Results from {\tt CMBACT} using 2000 network realisations are shown in solid black, and from our code using the first 16 (dotted red), 32 (dashed blue), 64 (dot-dash green), 128 (dot-dot-dash magenta) and 256 eigenmodes (long-dashed cyan). We use string parameters of $G\mu = 2 \times 10^{-7}$, $v=0.65$, $\xi=0.13$ and $\alpha=1.9$. } 
\end{figure*}

The Einstein-Boltzmann solver we use is the publicly available {\tt CAMB} code~\cite{Lewis:1999bs}. This has several advantages over the {\tt CMBFAST}~\cite{Seljak:1996is} code which {\tt CMBACT} is based on. It has been tuned for efficiency and accuracy, and also uses the OpenMP parallel framework. Our implementation is based on the January 2012 version of the code. 

The string eigenvector sources were next incorporated.  For scalars we also require the longitudinal component of the velocity perturbation, $\Theta_D$. This can be derived from the known components by solving the differential equation~\cite{Albrecht:1997mz}
\beq
\dot{\Theta}_D = - 2 \mathcal{H} \Theta_D - \frac{k^2}{3} \left[ \frac{\Theta_D - \dot{\Theta}_{00} }{\mathcal{H}} - \Theta_{00} + 2 \Theta^S \right]\,,
\eeq
where overdots now represent differentiation with respect to conformal time and $\mathcal{H} = \dot{a}/a$.

The CMB power spectra do not require any initial spectrum to be specified, since the sources are active, and are given by 
\beq
C^{i \, (I)}_{\ell} = \frac{2}{\pi} \int k^2 dk \, \Delta^{i \, (I)}_{\ell} (k,\tau_0) \Delta^{i \, (I)}_{\ell} (k,\tau_0)\,,
\eeq
where $I=S,V,T$ for the scalar, vector and tensor contributions, and $\Delta^i_{\ell} (k,\tau_0)$  are the temperature/polarisation transfer functions for each mode. These transfer functions can be found in detail, for example, in Ref.~\cite{Hu:1997hp}.

For our analysis we use the mean \WMAP~7-year cosmological parameters~\cite{Komatsu:2010fb} -- that is a baryon density $\Omega_b h^2 = 0.02249$, cold dark matter density $\Omega_b h^2 = 0.1120$, Hubble parameter $H_0 = 70.4 \, {\rm km} \, {\rm s}^{-1} \, {\rm Mpc}^{-1}$, and an optical depth to reionisation $\tau=0.088$. We switch  off lensing of the CMB in order to directly compare results with {\tt CMBACT}.  The same ``default'' set of string parameters of $v=0.65$, $\xi=0.13$ and $\alpha = 1.9$ are used, again turning off the time evolution of these and setting the decay parameter to $L_f=0.99$ in {\tt CMBACT}. We fix the string tension to $G\mu = 2 \times 10^{-7}$, a limit which is allowed, for the default parameters, by current CMB data~\cite{Battye:2010xz}. 

\begin{figure*}
\centering
\mbox{\resizebox{0.8\textwidth}{!}{\includegraphics[angle=0]{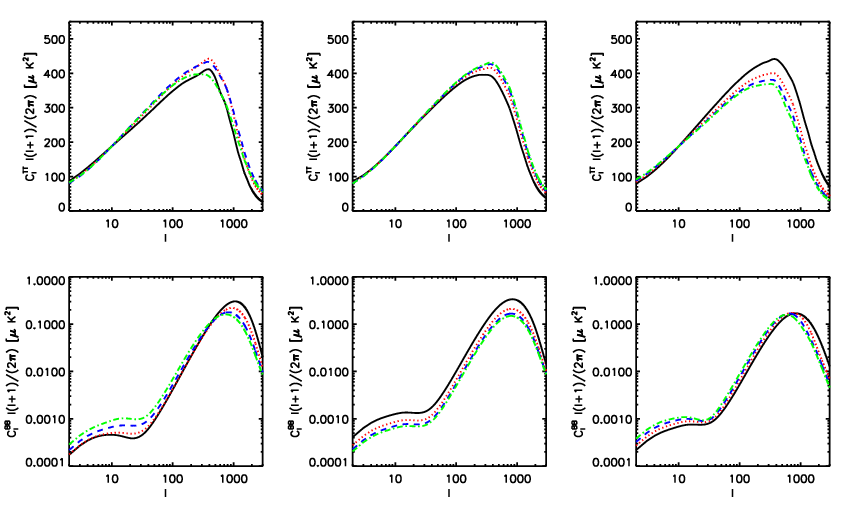}}}
\caption{\label{fig:uetc_cls_ttbb} String parameter dependence of the {\em total} TT (top) and BB (bottom) spectra. Results are derived using 256 eigenmodes and a weighting of $\gamma=0.25$.  In the left panels we set $\xi=0.13$, $\alpha=1.9$ and vary $v$ as 0.2 (solid black), 0.4 (dotted red), 0.6 (dashed blue), 0.8 (dot-dash green). In the middle we set $\xi=0.13$, $v=0.65$ and vary $\alpha$ as 1.0 (solid black), 1.5 (dotted red), 2.0 (dashed blue), 2.5 (dot-dash green). On the right we set $v=0.65$, $\alpha=1.9$ and vary $\xi$ as 0.1 (solid black), 0.2 (dotted red), 0.3 (dashed blue), 0.4 (dot-dash green).  In each case $G\mu = 2 \times 10^{-7}$.} 
\end{figure*}

As in the {\tt CMBACT} implementation, we found it necessary to adjust internal accuracy parameters of {\tt CAMB} to achieve the desired accuracy in the power spectrum. We varied the number of terms in the Boltzmann hierarchy and the $\ell$ mode spacing to compute the spectrum, but the only adjustment we found necessary was to increase the number of $k$ modes for the CMB source functions. For the aficionados, sufficient accuracy was achieved by setting {\tt high\_accuracy\_default = T} and {\tt accuracy\_boost = 1.5}.

The TT, EE, BB and TE power spectra are shown in Fig.~\ref{fig:uetc_cls}, computing the UETC's on a 1024 x 1024 grid in the range $k \tau=[10^{-4}, 10^4]$, with logarithmic grid spacing, and a weighting factor of $\gamma=0.25$. The range and grid size of the UETC were chosen so that the spectrum is insensitive to increasing the range or number of grid points. The weighting factor has been chosen to improve the convergence of the spectrum for a finite number of modes. 

We plot spectra truncating the eigenmode sum at 16, 32, 64, 128 and 256 modes, and also show the results from {\tt CMBACT} using 2000 network realisations. The convergence rate depends on the particular spectrum of interest. In general, polarisation spectra converge more quickly since (aside from the reionisation induced signal) they are sourced only at last scattering. This means that a smaller range of scales  (and hence eigenmodes) contribute to the source function, and the weighting factor is chosen to preferentially reconstruct these modes. The vector B-mode spectrum, for example, converges in as little as 32 eigenmodes. The scalar temperature signal, on the other hand, is sourced both at last scattering and along the line-of-sight through the Integrated Sachs Wolfe effect. This means a wider range of eigenmodes contribute to the spectrum.

Plotting the difference of each spectra we find that $1\%$ level accuracy in the temperature power spectrum can be achieved with 256 modes. On a 3.1 GHz Intel Xeon CPU with 8 threads, our code runs  in $\sim 15$ minutes. In comparison, around 2000 network realisations are required for {\tt CMBACT} to achieve the same accuracy -- on the same CPU, running with only the single thread that is possible, the computation  time is $\sim 30$ hours. If accuracy is only required in the temperature power spectra at the $10 \%$  level, such that 64 eigenmodes are required, our  code takes $\sim 2.5$ minutes with 16 threads. Given that strings are constrained to $< 10 \%$ of the total temperature anisotropy  of the CMB, this level of accuracy is tolerable for parameter estimation studies. For the first time it may be possible to explore the string parameter space using Markov-Chain-Monte-Carlo (MCMC) methods.

Generally, the results from our code and {\tt CMBACT} agree very well, considering the two different implementations. The only differences of note are a slight excess in power in the {\tt CMBACT} scalar and tensor temperature fluctuations near the peak of the spectra, a slight excess in the low $\ell$ tensor E-mode polarisation, and a small difference in the tensor TE spectrum at $\ell \sim 800$. We have been unable to reconcile these small differences by changing details of the UETC grid, accuracy or increasing the number of modes. 


\subsection{Parameter dependence}

With our code in hand, we can now begin to explore the string model space. Some of these parameter dependences have also been considered in Refs.~\cite{Albrecht:1997mz,cmbact,Pogosian:2007gi}. The default model is specified by $v=0.65$, $\xi=0.13$ and $\alpha = 1.9$. In each case we fix two of the parameters while varying the other. We again compute the UETC's on a 1024 x 1024 grid in the range $k \tau=[10^{-4}, 10^4]$ and a weighting of $\gamma=0.25$, truncating the sum at 256 modes. 

In Fig.~\ref{fig:uetc_cls_ttbb} we show the total TT and BB power, normalising the value of $G\mu$ such that the TT  power at $\ell=10$ is the same as the default model. Given the default model has a tension $G\mu_0 = 2 \times 10^{-7}$, the ratio of $\mu$ required for the same $C_{10}$ is $\mu/\mu_0 = [1.99, 1.57, 0.11, 0.65]$, for $v=0.2$, 0.4, 0.6 and 0.8 respectively. For $\alpha$ the ratio is  $\mu/\mu_0 = [1.50, 1.20, 0.95, 0.78]$ for $\alpha=1.0$, 1.5, 2.0 and 2.5. Finally, for $\xi$ the ratio is $\mu/\mu_0 = [0.84, 0.33, 1.79, 2.25]$ for $\xi=0.1$, 0.2, 0.3 and 0.4. One can see that $G \mu$ has a strong dependence on the values of the string parameters. 
 
We leave a detailed exposition of the dependence of the power spectra on the string parameters to future work, but some of the broad features can be understood by considering the analytic form of the UETC given in Appendix~\ref{app:int}. The low $\ell$ amplitude is determined by the small $x$ limit (the $B$ coefficients). For scalars, this is complicated by the fact that density, velocity and anisotropic stress all act as sources. For vectors and tensors, however, the low $\ell$ dependence of both temperature and polarisation (ignoring the reionisation induced polarisation signal) scales according to
\beq
C_{\ell}^{(V, T)} \propto  (G\mu)^2  \frac{ \left[ 1+ v^2 (\alpha^2-2) +v^4 (1-\alpha^2 + \alpha^4)  \right]}{ \xi  \alpha^2 (1-v^2)}\,.
\eeq
 As far as we are aware this explicit parameter dependence has not been written down before. We have tested this with the actual $C_{\ell}$'s from {\tt CAMB}, and find the agreement for polarisation from $\ell \sim 40$ to several hundred (where the spectrum begins to turn over) to be extremely accurate. The relationship is less accurate for temperature (within a factor of two for the range of parameters we consider), since it is also sourced by smaller scales along the line-of-sight. 
  
 Other features of the spectrum can also begin to be understood using the analytic form of the UETC. Changing the velocity, for example, has a larger effect on the B-mode peak position than temperature. The scalar contribution to the latter is dominated by the density fluctuations near the position of the peak. Examining the UETC coefficients, one can see that velocity has {\em no effect} on the amplitude of these coefficients. In particular, along the diagonal the velocity has no effect on the scale at which the UETC turns over, as can be seen from the form of the ETC (the $C$ coefficient). The scale at which the vector anisotropic stress UETC turns over, however, is affected by the velocity. Expanding out the coefficients one can see that a higher velocity causes the UETC to turn over on larger scales, which is consistent with the shift in the B-mode peak position. 
  
 Note that this exploration of parameter dependencies by varying one network parameter at 
a time is strictly not allowed, as the network parameters are related within a given scaling solution, as given by Eqns.~(\ref{keqn}, \ref{vosscaling1}, \ref{vosscaling2}). Here we have used this as a phenomenological tool for singling out the dependencies of the UETC on the corresponding physical variables, which can help develop an intuition of the physics involved~\cite{inprep}. 
      

\section{Conclusions}

 The main achievement of this paper has been to derive} analytic expressions  (see Appendix~\ref{app:int}) for the UETC's of cosmic strings in the USM. These expressions agree extremely well with simulating an ensemble of source histories. Our formalism makes the physical connection with CMB anisotropies more transparent, as UETC's can be obtained very efficiently for different assumptions regarding the properties of the string network. From this we have written a new code to compute CMB power spectra.   We believe that our implementation improves that of the existing {\tt CMBACT} code in three main ways:
\begin{enumerate}
\item The UETC's are calculated analytically. While {\tt CMBACT} does not directly compute UETC's, there is computational overhead associated with creating an ensemble of source histories. 
\item Once the UETC's have been calculated they can be diagonalised, with the eigenvectors acting as sources in an Einstein-Boltzmann code. A much smaller number of eigenmodes, as opposed to averaging over a number of source histories, are required to achieve the same accuracy in the CMB power spectrum.  
\item Our implementation uses the {\tt CAMB} Einstein-Boltzmann code, which has several advantages over {\tt CMBFAST}, which {\tt CMBACT} is based on. {\tt CAMB} has been optimised for efficiency and accuracy, is more modular, and supports the OpenMP framework. Secondly, {\tt CAMB} supports the latest features such as CMB lensing and bispectrum estimation, and is readily compatible with the Markov-Chain-Monte-Carlo (MCMC) package {\tt CosmoMC}~\cite{Lewis:2002ah}.  
\end{enumerate}
The resulting code can compute spectra, whose temperature power spectra are accurate at the $10 \%$  level, in only a few minutes on a multi-threaded CPU. 

The dramatic improvement in computational time achieved with our approach
now allows us to perform comprehensive scans over the string network parameter 
space. In particular, the compatibility of our {\tt CAMB} implementation with {\tt CosmoMC} 
allows us to perform cosmological parameter fitting for inflation+strings scenarios, 
marginalising over string network parameters. This will be the subject of a forthcoming 
publication.  Our methods can also be generalised to more complicated networks of 
cosmic superstrings, comprising string segments with different tensions that join together 
in Y-shaped junctions~\cite{inprep}.  This will allow us to efficiently 
probe a wide range of well-motivated inflationary models in string theory, constraining 
fundamental parameters~\cite{ACMPPS}. 


\section*{Acknowledgments} 
This research was supported by STFC.  The work of AA was supported by the 
Marie Curie grant FP7-PEOPLE-2010-IEF-274326 at the University of Nottingham.
EJC would also like to thank the  Royal Society and Leverhulme Trust for financial support. We thank Richard Battye for very useful discussions and the referee for checking our UETC coefficients.    


\appendix

\section{Integral identities and amplitudes} \label{app:int}

This appendix details integral identities and amplitudes used in deriving the UETC's. Note there is actually a divergent component in both the $I_1 (x,\rho) $ and $I_4 (x,\rho) $ integrals, but these cancel as we always take the combination $I_i (x_-, \rho) - I_i (x_+, \rho) $. Therefore in the following relations the divergent component is discarded. 

\begin{widetext}
\begin{equation} \label{eqn:j0_series}
I_1 (x,\rho) = \frac{1}{2} \int_0^{\pi} d \theta \sin \theta \cos(x \cos \theta)  J_0 (\rho \sin \theta) \sec^2 \theta =  \sum_{c=0}^{\infty} \frac{1}{c!}  \frac{\rho}{(2c-1)} \left(-\frac{x^2}{2 \rho} \right)^c j_{c-1} (\rho)\,,
\end{equation}
where $j_n(x)$ is the spherical Bessel function.

\begin{equation}
I_2 (x,\rho) = \frac{1}{2} \int_0^{\pi} d \theta \sin \theta \cos(x \cos \theta) J_0 (\rho \sin \theta) =  \left( \frac{\sin \sqrt{\rho^2 +x^2} }{\sqrt{\rho^2 +x^2}} \right)  \,, 
\end{equation}

\begin{equation}
I_3 (x,\rho) = \frac{1}{2} \int_0^{\pi} d \theta \sin^3 \theta \cos(x \cos \theta) J_0 (\rho \sin \theta) =  \left[ 1 + \frac{\partial^2}{\partial x^2} \right] \left( \frac{\sin \sqrt{\rho^2 +x^2} }{\sqrt{\rho^2 +x^2}} \right)  \,, 
\end{equation}

\begin{equation} \label{eqn:j1_series}
I_4(x,\rho) = \frac{1}{2} \int_0^{\pi} d \theta \sin \theta \cos(x \cos \theta)   \frac{J_1 (\rho \sin \theta)}{\rho \sin \theta} \sec^2 \theta =   \frac{\cos x}{\rho^2} -  \sum_{c=0}^{\infty} \frac{1}{c!} \frac{1}{(2c-1)}\left(-\frac{x^2}{2 \rho} \right)^c j_{c-2} (\rho)  \,,
\end{equation}

\begin{equation}
I_5(x,\rho) = \frac{1}{2}  \int_0^{\pi} d \theta \sin \theta \cos(x \cos \theta) \frac{J_1 (\rho \sin \theta)}{\rho \sin \theta} = \frac{1}{\rho^2} \left[\cos x- \cos \sqrt{\rho^2 +x^2}  \right]\,.
\end{equation}

\begin{equation}
I_6(x,\rho) = \frac{1}{2} \int_0^{\pi} d \theta \sin^3 \theta \cos(x \cos \theta) \frac{J_1 (\rho \sin \theta)}{\rho \sin \theta} = -\frac{1}{\rho^2 + x^2} \left[1+ \frac{1}{x} \frac{\partial}{\partial x} \right] \cos \sqrt{\rho^2 +x^2} \,.
\end{equation}
\end{widetext}

We now give amplitudes of the integral components and the $B$ and $C$ coefficients for each UETC. In the following ${\rm Si}[x]= \int_0^x {\rm sinc} \, x' \, dx'$ and ${\rm sinc} \, x = \sin x /x$.

\bigskip

$\langle \Theta_{00}(k, \tau_1) \Theta_{00}(k,\tau_2) \rangle:$
\begin{align} 
A_1 &=& 2 \alpha^2\,, \nonumber \\ \nonumber
A_2 &=& 0 \,, \\ \nonumber
A_3 &=& 0 \,, \\ \nonumber
A_4 &=& 0 \,, \\ \nonumber
A_5 &=& 0 \,, \\ \nonumber
A_6 &=& 0 \,, \\ \nonumber
B &=& \alpha^2 x_1 x_2\,, \\ \nonumber
C &=& 2 \alpha^2  (-1+\cos x+x\, \text{Si}[x]) \,, 
\end{align}

$\langle \Theta_{00}(k, \tau_1) \Theta^S(k,\tau_2) \rangle:$ 
\begin{align} 
A_1 &=& 1 + v^2 (2 \alpha^2-1) \,, \nonumber \\ \nonumber
A_2 &=& - 3 (1+v^2 (\alpha^2-1)) \,, \\ \nonumber
A_3 &=& 0  \,, \\ \nonumber
A_4 &=& - 3 v^2 \alpha^2  \,, \\ \nonumber
A_5 &=& 3 v^2 \alpha^2  \,, \\ \nonumber
A_6 &=& 0 \,, \\ \nonumber
B &=&   \frac{ x_1 x_2 (x_1^2 +x_2^2 ) (2+v^2(\alpha^2-2))}{360}\,, \\ \nonumber
&-&   \frac{ x_1 x_2 \rho^2 (1+v^2 (2 \alpha^2-1)) }{30}\,, \\ \nonumber
C &=& \frac{ (2+v^2(\alpha^2-2)) (-4+\cos x+3 \, {\rm sinc}\, x + x\, \text{Si}[x])}{2} \,, 
\end{align}

$\langle \Theta^S(k, \tau_1) \Theta^S(k,\tau_2) \rangle:$
\begin{align} 
A_1 &=& \frac{-27 v^4 \alpha^4 + \rho^2 (1+v^2(2 \alpha^2-1))^2}{2 \alpha^2 \rho^2} \,, \nonumber \\ \nonumber
A_2 &=& \frac{3(9 v^4 \alpha^4 + \rho^2 (1-2v^2 - v^4(\alpha^4-1)))}{2 \alpha^2 \rho^2}  \,, \\ \nonumber
A_3 &=& - \frac{9 (1+v^2(\alpha^2-1))^2}{2 \alpha^2} \,, \\ \nonumber
A_4 &=& 3 v^2 \left(v^2 \left(1+ \alpha^2 \left(\frac{9}{\rho^2}-2 \right) \right)-1 \right)   \,, \\ \nonumber
A_5 &=& -3 v^2 \left(v^2 \left(1+ \alpha^2 \left(\frac{9}{\rho^2}-2 \right) \right)-1 \right)    \,, \\ \nonumber
A_6 &=& 9 v^2(1 + v^2 (\alpha^2-1))\,, \\ \nonumber
B &=& \frac{ x_1 x_2 \left(1+ v^2 (\alpha^2-2) +v^4 (1-\alpha^2 +  \alpha^4 )\right) }{5 \alpha^2}  \,, \\ \nonumber
C &=& \frac{\cos x \left[  (x^2-18)+v^2 (x^2-18)(\alpha^2-2) \right]}{2 x^2 \alpha^2}  \\ \nonumber &+& \frac{ \cos  x \, v^4 \left[x^2 \left(1-\alpha^2 + \frac{11 \alpha^4}{8} \right) -18 \left(1-\alpha^2+\frac{3 \alpha^4}{8} \right) \right]}{2 x^2 \alpha^2}  \\ \nonumber &-& \frac{2 \left[1+v^2 (\alpha^2-2)+v^4 (1-\alpha^2 + \alpha^4) \right]}{\alpha^2}  \\ \nonumber &+& \frac{x \, {\rm Si}[x] \left[1+v^2 (\alpha^2-2)+v^4 \left(1-\alpha^2 + \frac{11 \alpha^4}{8} \right) \right]}{2 \alpha^2}  \\ \nonumber &+& \frac{3 \, {\rm sinc} \, x \left[(6-x^2)+v^2 (6-x^2)(\alpha^2-2) \right] }{2 x^2 \alpha^2}  \\ \nonumber &+&  \frac{ 3 \, {\rm sinc} \, x  \, v^4 \left[6\left(1-\alpha^2 + \frac{3 \alpha^4}{8} \right)-x^2 \left(1-\alpha^2 - \frac{\alpha^4}{8}  \right) \right] }{2 x^2 \alpha^2}\,,
\end{align}

$\langle \Theta^T(k, \tau_1) \Theta^T(k,\tau_2) \rangle$:
\begin{align} 
A_1 &=&  \frac{\rho^2 (v^2-1)^2-3\alpha^4 v^4}{4 \alpha^2 \rho^2}\,, \nonumber \\ \nonumber
A_2 &=&  \frac{\rho^2 (-1 + 2v^2 + v^4 (\alpha^4-1))+3 \alpha^4 v^4}{4 \alpha^2 \rho^2}\,, \\ \nonumber
A_3 &=& - \frac{(1+v^2(\alpha^2-1))^2}{4 \alpha^2}\,, \\ \nonumber
A_4 &=& \frac{v^2}{2} \left(1+v^2 \left(\frac{3\alpha^2}{\rho^2}-1 \right) \right)  \,, \\ \nonumber
A_5 &=&  -\frac{v^2}{2} \left(1+v^2 \left(\frac{3\alpha^2}{\rho^2}-1 \right) \right) \,, \\ \nonumber
A_6 &=& \frac{v^2}{2} (1+v^2 (\alpha^2-1)) \,, \\ \nonumber 
B &=&  \frac{ x_1 x_2 \left(1+ v^2 (\alpha^2-2) +v^4 (1-\alpha^2 + \alpha^4 )\right) }{15 \alpha^2}\,, \\ \nonumber
C  &=&  \frac{   {\rm sinc}\, x \left[1 + v^2 (\alpha^2-2) + v^4 \left(1-\alpha^2+\frac{3  \alpha^4}{8} \right) \right] }{2 x^2 \alpha^2}   \\ \nonumber &+& \frac{ {\rm sinc}\, x \left[1+v^2 (\alpha^2-2)+v^4 (1-\alpha^2 - \frac{5 \alpha^4}{8}) \right] }{4 \alpha^2}   \\ \nonumber 
&-& \frac{2 (1-v^2)(1+v^2(\alpha^2-1)  ) }{3  \alpha^2}  \\ \nonumber
&+& \left\{\frac{ \left[ (1-\frac{2}{x^2}) \cos x + x \, \text{Si}[x] \right] }{4 \alpha^2} \right. \\ \nonumber 
& & \times \left. \left[1+v^2(\alpha^2-2)+v^4 \left(1-\alpha^2+\frac{3  \alpha^4}{8} \right) \right] \right\}\,, 
\end{align}

$\langle \Theta^V(k, \tau_1) \Theta^V(k,\tau_2) \rangle$:
\begin{align} 
A_1 &=& \frac{3 v^4 \alpha^2}{\rho^2} \,, \nonumber \\ \nonumber
A_2 &=& - \frac{3 v^4 \alpha^2}{\rho^2} \,, \\ \nonumber
A_3 &=& \frac{1}{\alpha^2} \left(1+v^2 (\alpha^2-1) \right)^2 \,, \\ \nonumber
A_4 &=& - v^4 \alpha^2 \left(\frac{6}{\rho^2}-1 \right)  \,, \\ \nonumber
A_5 &=& v^4 \alpha^2 \left(\frac{6}{\rho^2}-1 \right)  \,, \\ \nonumber
A_6 &=& - 2v^2 \left(1+v^2 (\alpha^2-1) \right)\,, \\ \nonumber 
B &=& \frac{ x_1 x_2 \left(1+ v^2 (\alpha^2-2) +v^4 (1-\alpha^2 +   \alpha^4  )\right) }{15 \alpha^2}\,, \\ \nonumber
C &=& \frac{ v^4 \alpha ^2 \left(-2+\cos x+{\rm sinc}\, x+x\, \text{Si}[x]\right) }{8}   \\ \nonumber
&+& \left\{ \frac{ 2 \left[1+ v^2 \left(\alpha ^2-2\right)+v^4 \left(1- \alpha ^2+\frac{3\alpha ^4}{8} \right)\right] }{x^3 \alpha ^2}  \right. \\ \nonumber && \times \left. \left[\frac{x^3}{3}+ x \,\cos x-\sin x\right] \right\}\,,
\end{align}


\begin{thebibliography}{99}

\bibitem{Smoot:1992td}
  G.~F.~Smoot, C.~L.~Bennett, A.~Kogut, E.~L.~Wright, J.~Aymon, N.~W.~Boggess, E.~S.~Cheng and G.~De Amici {\it et al.},
  Astrophys.\ J.\  {\bf 396} (1992) L1.
  
\bibitem{Spergel:2003cb}
  D.~N.~Spergel {\it et al.}  [WMAP Collaboration],
  Astrophys.\ J.\ Suppl.\  {\bf 148} (2003) 175
  [astro-ph/0302209].

\bibitem{Linde:1993cn}
  A.~D.~Linde,
  Phys.\ Rev.\ D {\bf 49} (1994) 748
  [astro-ph/9307002].

\bibitem{Copeland:1994vg}
  E.~J.~Copeland, A.~R.~Liddle, D.~H.~Lyth, E.~D.~Stewart and D.~Wands,
  Phys.\ Rev.\ D {\bf 49} (1994) 6410
  [astro-ph/9401011].

\bibitem{Dvali:1994ms}
  G.~R.~Dvali, Q.~Shafi and R.~K.~Schaefer,
  Phys.\ Rev.\ Lett.\  {\bf 73} (1994) 1886
  [hep-ph/9406319].
  
\bibitem{Dvali:1998pa}
  G.~R.~Dvali and S.~H.~H.~Tye,
  Phys.\ Lett.\ B {\bf 450} (1999) 72
  [hep-ph/9812483].

\bibitem{BMNQRZ}
  C.~P.~Burgess, M.~Majumdar, D.~Nolte, F.~Quevedo, G.~Rajesh and R.~-J.~Zhang,
  JHEP {\bf 0107} (2001) 047
  [hep-th/0105204].

\bibitem{KKLMMT}
  S.~Kachru, R.~Kallosh, A.~D.~Linde, J.~M.~Maldacena, L.~P.~McAllister and S.~P.~Trivedi,
  JCAP {\bf 0310} (2003) 013
  [hep-th/0308055].

\bibitem{BDKMcAll}
  D.~Baumann, A.~Dymarsky, I.~R.~Klebanov and L.~McAllister,
  JCAP {\bf 0801} (2008) 024
  [arXiv:0706.0360 [hep-th]].

\bibitem{Jeannerot:2003qv}
  R.~Jeannerot, J.~Rocher and M.~Sakellariadou,
  Phys.\ Rev.\ D {\bf 68} (2003) 103514
  [hep-ph/0308134].

\bibitem{SarTye}
  S.~Sarangi and S.~H.~H.~Tye,
  Phys.\ Lett.\ B {\bf 536} (2002) 185
  [hep-th/0204074].

\bibitem{CMP}
  E.~J.~Copeland, R.~C.~Myers and J.~Polchinski,
  JHEP {\bf 0406} (2004) 013
  [hep-th/0312067].

\bibitem{Polch_Intro}
  J.~Polchinski,
  arXiv:hep-th/0412244.

\bibitem{CopKib}
  E.~J.~Copeland and T.~W.~B.~Kibble,
  Proc.\ Roy.\ Soc.\ Lond.\ A {\bf 466} (2010) 623
  [arXiv:0911.1345 [hep-th]].

\bibitem{DvalVil}
  G.~Dvali and A.~Vilenkin,
  JCAP {\bf 0403} (2004) 010
  [hep-th/0312007].

\bibitem{ACMPPS}
  A.~Avgoustidis, E.~J.~Copeland, A.~Moss, L.~Pogosian, A.~Pourtsidou and D.~A.~Steer,
  Phys.\ Rev.\ Lett.\  {\bf 107} (2011) 121301
  [arXiv:1105.6198 [astro-ph.CO]].

\bibitem{CopPogVach}
  E.~J.~Copeland, L.~Pogosian and T.~Vachaspati,
  Class.\ Quant.\ Grav.\  {\bf 28} (2011) 204009
  [arXiv:1105.0207 [hep-th]].

\bibitem{Hindm_Rev}
  M.~Hindmarsh,
  Prog.\ Theor.\ Phys.\ Suppl.\  {\bf 190} (2011) 197
  [arXiv:1106.0391 [astro-ph.CO]].

\bibitem{Kaiser:1984iv}
  N.~Kaiser and A.~Stebbins,
  Nature {\bf 310} (1984) 391.

\bibitem{Ringeval:2010ca}
  C.~Ringeval,
  Adv.\ Astron.\  {\bf 2010} (2010) 380507
  [arXiv:1005.4842 [astro-ph.CO]].

\bibitem{Battye:2010xz}
  R.~Battye and A.~Moss,
  Phys.\ Rev.\ D {\bf 82} (2010) 023521
  [arXiv:1005.0479 [astro-ph.CO]].

\bibitem{Urrestilla:2011gr}
  J.~Urrestilla, N.~Bevis, M.~Hindmarsh and M.~Kunz,
  JCAP {\bf 1112} (2011) 021
  [arXiv:1108.2730 [astro-ph.CO]].
  
\bibitem{Dvorkin:2011aj}
  C.~Dvorkin, M.~Wyman and W.~Hu,
  Phys.\ Rev.\ D {\bf 84} (2011) 123519
  [arXiv:1109.4947 [astro-ph.CO]].

\bibitem{TurPenSelj}
  N.~Turok, U.~-L.~Pen and U.~Seljak,
  Phys.\ Rev.\ D {\bf 58} (1998) 023506
  [astro-ph/9706250].

\bibitem{Albrecht:1997mz}
A.~Albrecht, R.~A.~Battye and J.~Robinson, 
 Phys.\ Rev.\  D {\bf 59} (1999) 023508 [arXiv:astro-ph/9711121].

\bibitem{cmbact}
L.~Pogosian and T.~Vachaspati, 
Phys.\ Rev.\  D {\bf 60} (1999) 083504 [arXiv:astro-ph/9903361].

\bibitem{book}
     A.~Vilenkin, and E.~P.~S.~Shellard,
     {\it Cosmic Strings and Other Topological Defects},
    (Cambridge University Press, 1994)

\bibitem{Hindmarsh:1994re}
  M.~B.~Hindmarsh and T.~W.~B.~Kibble,
  Rept.\ Prog.\ Phys.\  {\bf 58} (1995) 477
  [arXiv:hep-ph/9411342].
 
\bibitem{Vincent:1997cx}
  G.~Vincent, N.~D.~Antunes and M.~Hindmarsh,
  Phys.\ Rev.\ Lett.\  {\bf 80} (1998) 2277
  [hep-ph/9708427].
  
\bibitem{Moore:2001px}
  J.~N.~Moore, E.~P.~S.~Shellard and C.~J.~A.~P.~Martins,
  Phys.\ Rev.\ D {\bf 65} (2002) 023503
  [hep-ph/0107171].
  
\bibitem{Bevis:2006mj}
  N.~Bevis, M.~Hindmarsh, M.~Kunz and J.~Urrestilla,
  Phys.\ Rev.\ D {\bf 75} (2007) 065015
  [astro-ph/0605018].
 
\bibitem{Hindmarsh:2008dw}
  M.~Hindmarsh, S.~Stuckey and N.~Bevis,
  Phys.\ Rev.\ D {\bf 79} (2009) 123504
  [arXiv:0812.1929 [hep-th]].
 
\bibitem{Shell_Recon}
  E.~P.~S.~Shellard,
  Nucl.\ Phys.\ B {\bf 283} (1987) 624.
 
\bibitem{Matzner}
  R.~A.~Matzner and J.~Mccracken,
  IN *NEW HAVEN 1988, PROCEEDINGS, COSMIC STRINGS* 32-41.
 
\bibitem{Bennett:1989yp}
  D.~P.~Bennett and F.~R.~Bouchet,
  Phys.\ Rev.\  D {\bf 41} (1990) 2408.

\bibitem{Allen:1990tv}
  B.~Allen and E.~P.~S.~Shellard,
  Phys.\ Rev.\ Lett.\  {\bf 64} (1990) 119.
 
\bibitem{MartShell}
  C.~J.~A.~P.~Martins and E.~P.~S.~Shellard,
  Phys.\ Rev.\ D {\bf 73} (2006) 043515
  [astro-ph/0511792].
 
\bibitem{RingSakBouch}
  C.~Ringeval, M.~Sakellariadou and F.~Bouchet,
  JCAP {\bf 0702} (2007) 023
  [astro-ph/0511646].
 
\bibitem{B-POS}
  J.~J.~Blanco-Pillado, K.~D.~Olum and B.~Shlaer,
  Phys.\ Rev.\ D {\bf 83} (2011) 083514
  [arXiv:1101.5173 [astro-ph.CO]].
 
\bibitem{Kibble}
  T.~W.~B.~Kibble,
  Nucl.\ Phys.\  B {\bf 252} (1985) 227
  [Erratum-ibid.\  B {\bf 261} (1985) 750].
  
\bibitem{Austin:1993rg}
  D.~Austin, E.~J.~Copeland and T.~W.~B.~Kibble,
  Phys.\ Rev.\ D {\bf 48} (1993) 5594
  [hep-ph/9307325].
  
\bibitem{VOS}
  C.~J.~A.~Martins and E.~P.~S.~Shellard,
  Phys.\ Rev.\  D {\bf 54} (1996) 2535
  [arXiv:hep-ph/9602271].

\bibitem{VOSk}
  C.~J.~A.~Martins and E.~P.~S.~Shellard,
  Phys.\ Rev.\  D {\bf 65} (2002) 043514
  [arXiv:hep-ph/0003298].
  
\bibitem{Witten:1984eb}
  E.~Witten,
  Nucl.\ Phys.\ B {\bf 249} (1985) 557.

\bibitem{Oliveira:2012nj}
  M.~F.~Oliveira, A.~Avgoustidis and C.~J.~A.~P.~Martins,
  Phys.\ Rev.\ D {\bf 85} (2012) 083515
  [arXiv:1201.5064 [hep-ph]].
 
\bibitem{JackJoPolch}
  M.~G.~Jackson, N.~T.~Jones and J.~Polchinski,
  JHEP {\bf 0510} (2005) 013
  [arXiv:hep-th/0405229].

\bibitem{TWW}
  S.~H.~Tye, I.~Wasserman and M.~Wyman,
  Phys.\ Rev.\  D {\bf 71} (2005) 103508
  [Erratum-ibid.\  D {\bf 71} (2005) 129906]
  [arXiv:astro-ph/0503506].

\bibitem{NAVOS}
  A.~Avgoustidis and E.~P.~S.~Shellard,
  Phys.\ Rev.\  D {\bf 78} (2008) 103510
  [arXiv:0705.3395 [astro-ph]].
  
\bibitem{Avgoustidis:2009ke}
  A.~Avgoustidis and E.~J.~Copeland,
  Phys.\ Rev.\ D {\bf 81} (2010) 063517
  [arXiv:0912.4004 [hep-ph]].
 
\bibitem{wiggly}
  C.~J.~A.~P.~Martins,
  Astrophys.\ Space Sci.\  {\bf 261} (1999) 311.
 
\bibitem{Vilenkin_ss}
  A.~Vilenkin,
  Phys.\ Rev.\ D {\bf 41} (1990) 3038.
 
\bibitem{Carter_wiggly}
  B.~Carter,
  Phys.\ Rev.\ Lett.\  {\bf 74} (1995) 3098
  [hep-th/9411231].
 
\bibitem{Carter_super}
  B.~Carter and P.~Peter,
  Phys.\ Rev.\ D {\bf 52} (1995) 1744
  [hep-ph/9411425].
 
\bibitem{Battye:1997hu}
  R.~A.~Battye, J.~Robinson and A.~Albrecht,
  Phys.\ Rev.\ Lett.\  {\bf 80} (1998) 4847
  [astro-ph/9711336].
 
\bibitem{Pen:1997ae}
  U.~-L.~Pen, U.~Seljak and N.~Turok,
  Phys.\ Rev.\ Lett.\  {\bf 79} (1997) 1611
  [astro-ph/9704165].
 
\bibitem{Lewis:1999bs}
  A.~Lewis, A.~Challinor and A.~Lasenby,
  Astrophys.\ J.\  {\bf 538} (2000) 473
  [astro-ph/9911177].
 
\bibitem{Seljak:1996is}
  U.~Seljak and M.~Zaldarriaga,
  Astrophys.\ J.\  {\bf 469} (1996) 437
  [astro-ph/9603033].
 
\bibitem{Hu:1997hp}
  W.~Hu and M.~J.~White,
  Phys.\ Rev.\ D {\bf 56} (1997) 596
  [astro-ph/9702170].
  
\bibitem{Komatsu:2010fb}
  E.~Komatsu {\it et al.}  [WMAP Collaboration],
  Astrophys.\ J.\ Suppl.\  {\bf 192} (2011) 18
  [arXiv:1001.4538 [astro-ph.CO]].
 
\bibitem{Pogosian:2007gi}
  L.~Pogosian and M.~Wyman,
  Phys.\ Rev.\ D {\bf 77} (2008) 083509
  [arXiv:0711.0747 [astro-ph]].
 
\bibitem{inprep}
   A.~Avgoustidis, E.J.~Copeland, A.~Moss, and D.~Skliros,
   In preparation.   
 
\bibitem{Lewis:2002ah}
  A.~Lewis and S.~Bridle,
  Phys.\ Rev.\ D {\bf 66} (2002) 103511
  [astro-ph/0205436].
 
 
\end{thebibliography}
\end{document}